\documentclass{article}

 \usepackage[final]{neurips_2020}

\usepackage[utf8]{inputenc} 
\usepackage[T1]{fontenc}    
\usepackage{hyperref}       
\usepackage{url}            
\usepackage{booktabs}       
\usepackage{amsfonts}       
\usepackage{nicefrac}       
\usepackage{microtype}      
\usepackage{amsmath, amssymb}
\usepackage{enumitem}
\usepackage{graphicx}
\usepackage{subfigure}
\usepackage[export]{adjustbox}

\title{Applying Machine Learning to Crowd-sourced Data from Earthquake Detective}

\author{%
  Omkar Ranadive\ \\
  Department of Computer Science\\
  Northwestern University\\
  \texttt{omkar.ranadive@u.northwestern.edu} \\
   \And
   Suzan van der Lee \\
   Department of Earth and Planetary Sciences\\
   Northwestern University \\
   \texttt{suzan@northwestern.edu} \\
  \AND
  Vivian Tang \\
Department of Earth and Planetary Sciences\\
   Northwestern University \\
  \texttt{vivian@earth.northwestern.edu}
  \AND
 Kevin Chao \\
Data Science Initiative/NICO\\
Northwestern University \\
  \texttt{kchao@northwestern.edu} \\
}

\begin{document}

\maketitle

\begin{abstract}
Dynamically triggered earthquakes and tremor generate two classes of weak seismic signals whose detection, identification, and authentication traditionally call for laborious analyses. 
Machine learning (ML) has grown in recent years to be a powerful efficiency-boosting tool in geophysical analyses, including the detection of specific signals in time series. However, detecting weak signals that are buried in noise challenges ML algorithms, in part because ubiquitous training data is not always available. Under these circumstances, ML can be as ineffective as human experts are inefficient. 
At this intersection of effectiveness and efficiency, we leverage a third tool that has grown in popularity over the past decade: Citizen science. 
Citizen science project Earthquake Detective leverages the eyes and ears of volunteers to detect and classify weak signals in seismograms from potentially dynamically triggered (PDT) events. Here, we present the Earthquake Detective data set - A crowd-sourced set of labels on PDT earthquakes and tremor. 
We apply Machine Learning to classify these PDT seismic events and explore the challenges faced in segregating and classifying such weak signals. We confirm that with an image- and wavelet-based algorithm, machine learning can detect signals from small earthquakes. In addition, we report that our ML algorithm can also detect signals from PDT tremor, which has not been previously demonstrated. The citizen science data set of classifications and ML code are available online.\footnote{https://www.zooniverse.org/projects/vivitang/earthquake-detective }\footnote{https://github.com/Omkar-Ranadive/Earthquake-Detective}
\footnote{This paper is an updated version of the one presented at NeurIPS'20}

\end{abstract}
\section{Introduction}

Over the past five years, Machine Learning (ML) has progressively grown to be a popular tool in geophysical analyses. Much of this research demonstrates the impressive efficiencies that can be achieved by applying ML to tasks that are overwhelming for researchers from a data volume or dimensionality perspective while relatively straightforward in complexity or signal strength (\cite{10.1093/gji/ggaa444}; \cite{10.1093/gji/ggab083}; \cite{10.1093/gji/ggx238}; \cite{10.1093/gji/ggy385};). Few papers demonstrate ML's success in recognizing of low-amplitude signals in seismology \cite{rouet2017machine}. Here we leverage a crowd-sourced data set of weak seismic signals classified by citizen scientists \cite{tang2020citizen} in combination with a data set analyzed and labeled by experts (the authors) to establish a baseline for detecting different types of weak seismic signals. The ultimate goal of this work is to incorporate the ML algorithm in winnowing the data stream presented to citizen scientists and experts.

One class of weak seismic signals are seismograms from earthquakes with magnitudes below the magnitude of completion for typical earthquake catalogs. Yet these low-magnitude earthquakes belong to the same Gutenberg-Richter distribution \cite{gutenberg1954seismicity} as widely recorded earthquakes, including those that cause injuries, damage, and worse. Gutenberg and Richter's law (1954) states that for every magnitude (M) 8 earthquake that occurs, about 1 million M2 earthquakes occur.
Therefore, if we can detect the abundant low-magnitude earthquakes of the kind whose signals are often buried in background seismic noise, their analyses could provide insights into the occurrence, distribution, and physics of these \textit{and} the much sparser damaging earthquakes. 

Likewise, more recently investigated low-frequency earthquakes exist that represent slower slip between two blocks of rock than that during classical earthquakes, but nevertheless generate weak seismic signals that are often labeled as "tremor" \cite{obaratremor}, especially when many of such events occur quasi-simultaneously.

Weak signals from tremor and from low-magnitude local earthquakes are both abundant (\cite{rouet2018breaking}; \cite{hill2015dynamic}) and somewhat under-reported because they have been hard to detect. Past barriers to detection have included a sparse spatial distribution of seismic stations (instrumented with buried seismometers), and current barriers include the weakness and limited bandwidth of these weak seismic signals.  These detection challenges have inspired the application of machine-learning algorithms to large sets of seismic waveform data. Several of these ML algorithms have successfully been trained to detect signals from local earthquakes (e.g. \cite{ross2018generalized}; \cite{tang2020automating}) while others \cite{liu2019investigation} used ML to detect tremor signals. Using ML in the detection of signals from seismic activity is a rapidly growing field (e.g. \cite{bergen2019preface}; \cite{meier2019reliable}; Huang, 2019; \cite{li2018seismic}; \cite{riggelsen2014machine}; \cite{ruano2014seismic}; \cite{reynen2017supervised}; \cite{wiszniowski2021machine}).

Weak waveforms from abundant minor seismic events do not only add constraints to estimates of seismic risk, which are based on regional variations in the rate at and mode in which earthquakes occur, but also provide important information on where, when, and how they strike, allowing us to learn about the conditions under which earthquakes nucleate, occur, and interact. Therefore, the more we detect weak signals from minor seismic events, the more we learn about the physics and potential hazards of seismic slip. For example, we can learn about the dynamics of earthquake triggering by first detecting seismic events that occurred simultaneously with transient strain events, then determining the likelihood that these seismic events were triggered by the strain events, followed by examining the conditions under which such triggering does and does not occur (\cite{tang2021}).

Here, our interests lie in detecting a special sub-class of the multitude of minor seismic events, namely local earthquakes and tremor that could have been triggered by slowly-oscillating large-amplitude seismic surface waves from large-magnitude teleseismic earthquakes. Reporting and learning more about such Potentially Dynamically Triggered (PDT) events extends the spectrum of seismic slip data available for study and adds information about how fast and slow-slip earthquakes might nucleate.  Traditional ways for detecting signals form PDT events are 1) seismologists interactively examining seismograms and labeling detections after a range of signal inspections \cite{gomberg2008widespread}, and 2) seismologists developing and applying an automated detection algorithm to seismic waveform data while controlling the quality of the detections by tweaking the algorithm’s parameters and handling outliers separately (e.g. \cite{velasco2008global}; \cite{yun2021dyntripy}). 

In our quest to detect PDT seismic events we face a number of additional challenges:

\begin{enumerate}
\item The magnitudes (M) of PDT earthquakes are typically below the M of completeness of earthquake catalogs, hence their signals are weak and often buried in ambient seismic noise signals. Therefore, the database of template waveforms available for such low-M events is small at best. Signals from low-M events are not only lower in amplitude than those from higher-M events but also have narrower bandwidths, diminishing the efficacy of template matching methods. 
\item Unlike signals from dynamically triggered earthquakes, signals from dynamically triggered tremor have different waveforms than those from typical tremor on account of the former signals being modulated by the teleseismic surface waves that triggered them \cite{chao2012remote}). Therefore, a database of template signals is not available for training or other purposes, although a catalog has been started \cite{kano2018development}.

\item A signal from a PDT event can arrive at any time during the time window of surface wave passage, which is much longer in duration than the PDT event signal.  We have been considering up to 33 minutes of surface wave duration in labeling whether or not at least one local earthquake or tremor signal was recorded. 

\item Non-stationary noise signals often exceed or are comparable in amplitude and duration to the relatively weak signals of the PDT events we are interested in. 

\item Optimal and accurate detection of signals from PDT events requires a multi-scale, multi-band, multi-component interactive analysis that is labor-intensive. The formation of a large, labeled data set for training purposes is hence not straightforward. 

\end{enumerate}

\begin{figure}
  \includegraphics[width=\linewidth]{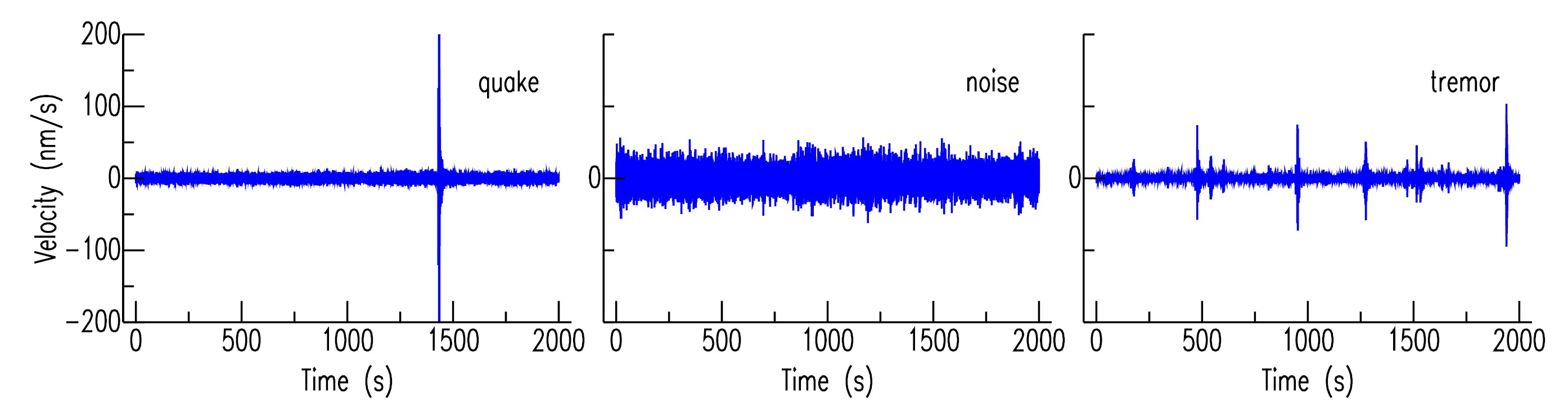}
  \caption{Example data plots shown to users on Earthquake Detective platform}
  \label{eqplots}
\end{figure}

\section{Earthquake Detective}
Earthquake Detective is a crowdsourcing platform where volunteers are shown a seismogram of vertical ground velocity vs. time (Figure \ref{eqplots}), along with a sonification of the signal, and are asked to classify the data as Earthquake/Tremor/Noise/None of the above. The platform currently has over 6000 volunteer scientists and over 130k classifications. \cite{tang2020citizen} present an analysis of how the volunteers and seismologists engage with the data. 

All raw input data are time series of recorded ground motion with durations of 2000 s, which are long enough to contain the time window needed by teleseismic surface waves to pass through. The raw data is demeaned, deconvolved with the instrument response to convert digital counts to physical units of ground velocity, band-pass filtered between 2-8 Hz with a 2-pole Butterworth filter, and resampled at 20 samples per second.

\section{Approach}
\subsection{Wavelet Scattering Transform}
The wavelet scattering transform decomposes a signal using a family of wavelets. This new representation is stable against deformations and is translation invariant. This family of wavelets does not need to be learned and hence, the features can be extracted without training which can then be passed on further to Machine Learning models \cite{oyallon2013generic}. The Scattering Transform has been successfully used by \cite{seydoux2020clustering} for clustering earthquakes in an unsupervised fashion. The scattering transform works by successively convolving wavelets with the signal and applying modulus non-linearity at each step. This can be shown as follows: 
\begin{equation}
  S_{x(t)} =  | | x * \psi_{\lambda1} |  * \psi_{\lambda 2}  | ... | * \psi_{\lambda m}| * \phi 
\end{equation}
where $S_{x(t)}$ is the set of scattering coefficients obtained at step m, $x(t)$ is the signal, $\psi_{\lambda_{m}}(t)$ denotes the set of wavelets at step m and $\phi(t)$ is a low-pass filter. 

In our experiments, we use the Kymatio library \cite{andreux2020kymatio} to perform the scattering transform and get a set of features which are passed on further to our supervised neural network model. 

\subsection{Convolution over 3-channel plots}
As the volunteers classify the data based on plots and associated audio records of vertical ground motion, we decided to try out a similar approach with our model. Along with the wavelet coefficients, we additionally provide the model with 3-channel plots (BHZ, BHE, BHN) as input. Image convolution is applied over the 3-channel plots and the resultant features are concatenated with the wavelet coefficients and then passed through a fully connected neural network (FCN). 

\begin{figure}
  \centering
   \includegraphics[width=\linewidth]{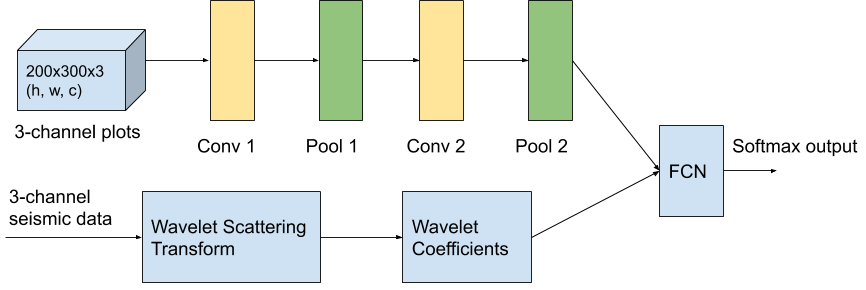}
  \caption{WavImg Model Architecture}
  \label{model}
\end{figure}

\section{Experiments}
For all experiments, we use the following hyper parameters: learning rate = 1e-5, batch size = 100, epochs = 300. We perform a 3-way classification between Earthquakes/Tremors/Noise. As the number of tremor samples is considerably less, we apply a weighted cross-entropy loss where the weights are calculated as follows: 
\begin{equation}
    w_{i} = N_{largest}/N_{i}
\end{equation}
where $w_{i}$ is the weight assigned to class i, $N_{largest}$ is the number of samples of the largest class and $N_{i}$ is the number of samples of class i.
For all experiments, we perform a 80-20 stratified train/test split of the data. We test two models: 

\begin{enumerate}
    \item \textbf{WavNet: }In this model, we perform a wavelet scattering transform on 3-channel seismic data (BHZ, BHN, BHE) and extract relevant features from it. These features (wavelet coefficients) are then passed on to a 2-layer fully connected network. 
    \item \textbf{WavImg: }This model combines the Wavelet Scattering Transform with 3-channel convolution over the image plots. The combined features are then passed on through a 2-layer fully connected network. (Figure \ref{model}) 
\end{enumerate}

\subsection{Training the Machine Learning model}

\textbf{Training on clean data: }
To test the efficacy of wavelet transform, we first run a simple experiment with WavNet. As an upper baseline, we first ran the experiment on clean data (data cleaned and filtered by our seismologists, refer Appendix A). The model converges to 95.2\% training and a 94.4\% testing accuracy.  

\textbf{Training on clean + gold users data: } 
Compared to the clean data, the data from the Earthquake Detective is difficult to segregate due to its low amplitude signals and larger time window ($\sim$33 mins). For this experiment, we consider data from gold users (Earthquake Detective data labeled by our experts) and combine it with the clean dataset from the previous experiment. When this data was trained on the WavNet, it gave a train accuracy of 75.4\% and a test accuracy of 74.9\%. Next, we trained the same data using the WavImg model. This model produces a 91.4\% train and 89.6\% test accuracy.  
 
\begin{figure}
 \centering 
  \includegraphics[width=\linewidth]{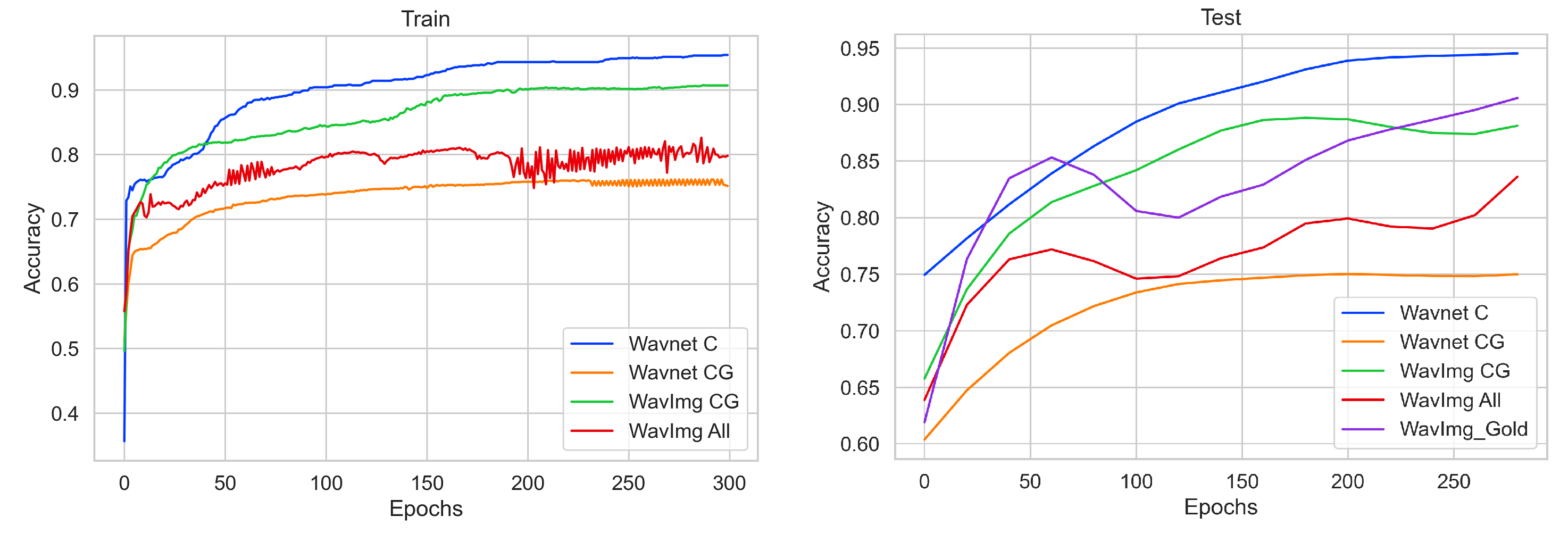}
  \caption{Model comparison chart. (C = Clean, CG = Clean+Gold, All = Clean+Gold+Volunteer)}
  \label{Comp_Chart} 
\end{figure}

\textbf{Training on clean + gold + volunteer's data: } 
Finally, we combine the previous data with data from two volunteers. For each volunteer, we calculated a reliability score which includes a precision, recall and f1 score  for each class. These scores were calculated by comparing volunteer's classification with gold-set labeled by our experts. 
To handle the unreliability introduced in labels, we add an additional gold-test set which consists of samples labeled from our gold users that were not used for training. WavImg produces a 80.1\% train accuracy, 83.6\% test accuracy and \textbf{90.4\%} gold-test accuracy. (Figure \ref{Comp_Chart})

\begin{table}
\caption{Comparison of model reliability with volunteers}
\centering
\begin{tabular}{llll}
\hline
 Metric & ML Model$^{*}$  & Volunteers & Volunteers (top 35\%) \\
\hline
 F1-Score (Earthquake) & 0.909 &  0.785 & 1.0 \\
 F1-Score (Noise)  & 0.914 & 0.738 & 0.975 \\
 F1-Score (Tremor) & 0.888 & 0.415 & 0.924  \\
\hline
\multicolumn{4}{l}{$^{*}$Here model refers to WavImg All (Clean + Gold + Volunteer)}
\end{tabular}
\end{table}

\subsection{Results and Analysis}
The WavNet model was able to perform extremely well on the clean data (95\% accuracy) which proves that wavelet scattering transform extracts relevant features which can then be trained using a simple 2-layer FCN. However, due to the greater complexity of Earthquake Detective data, WavNet by itself is insufficient. The WavImg model overcomes this problem by using information from 3-channel plots. One interesting case was the last experiment in which
there was variance in training due to label uncertainty (80\% accuracy) but the model still performed well on the gold-test set (90.4\% accuracy). (Figure \ref{Comp_Chart}) This shows that the model still ends up learning useful representations despite the uncertainty of the labels. For  more in-depth analysis over select misclassified samples refer to Appendix B.

\subsection{Model Comparison with Volunteers}
To see how reliable our model is in comparison to volunteers, we calculated the f1-score (reliability score) of the model using the gold-test for each class separately. The f1-score is calculated as follows: 
\begin{equation}
    F_{1} = (2*precision*recall)/(precision+recall)
\end{equation}
where precision is $TP/(TP+FP)$ and recall is $TP/(TP+FN)$, where TP is true positive, FP is false positive and FN is false negative. 

From table 1 we can see that the model is much more reliable for all classes when compared to all volunteers. However, we found that the volunteers become more reliable than the model in all cases, when top 35\% of them are selected. So this implies that while the model is better than an average volunteer, it is still not as good as the top volunteers.

\section{Conclusions}
The Earthquake Detective dataset is the first crowdsourced dataset for potentially dynamically triggered local earthquake and tremor signals. This is also the first time that potentially dynamically triggered tremor signals were used in and detected by ML.
Our experiments provide ML baselines for the data. We have only trained our models on a small subset of the 130k+ samples available. Also, from section 4.3 we saw that the top volunteers are better than the ML model. Therefore, when all of the data is considered, better techniques to incorporate reliability scores into the model will be required. 

We encourage researchers to 1) use this dataset as a catalog for potentially dynamically triggered seismic events, 2) augment the data and algorithms beyond the baseline, and 3) stream their data through Earthquake Detective for accelerating labeling of their data sets and/or for validating previously unlabeled ML results, by connecting with Earthquake Detective developers.

\section*{Broader Impact}
This work might facilitate work by research seismologists, seismic network operators, and seismic monitoring agencies and ultimately result in reducing risks posed by damaging earthquakes. 
Other impacts include the engagement of volunteer scientists, via Earthquake Detective, with informal STEM learning and analyzing authentic data.

\begin{ack}
This research was funded by the Integrated Data-Driven Discovery in Earth and Astrophysical Sciences (ID\textsuperscript{3}EAS) program under National Science Foundation grant NSF-NRT 1450006. We thank Josiah Evans and Blaine Rothrock for helpful discussions. A version of this paper was presented at the AI 4 Earth workshop at the NeurIPS 2020 conference; we thank the organizers for that opportunity to share and learn. This paper shares data generated via the zooniverse.org platform, development of which is funded by a Global Impact Award from Google and a grant from the Alfred P. Sloan Foundation. 
\end{ack}

\section*{}
\bibliographystyle{plainnat.bst}
\bibliography{sources.bib}

\begin{thebibliography}{29}
\providecommand{\natexlab}[1]{#1}
\providecommand{\url}[1]{\texttt{#1}}
\expandafter\ifx\csname urlstyle\endcsname\relax
  \providecommand{\doi}[1]{doi: #1}\else
  \providecommand{\doi}{doi: \begingroup \urlstyle{rm}\Url}\fi

\bibitem[Andreux et~al.(2020)Andreux, Angles, Exarchakis, Leonarduzzi,
  Rochette, Thiry, Zarka, Mallat, And{\'e}n, Belilovsky,
  et~al.]{andreux2020kymatio}
Mathieu Andreux, Tom{\'a}s Angles, Georgios Exarchakis, Roberto Leonarduzzi,
  Gaspar Rochette, Louis Thiry, John Zarka, St{\'e}phane Mallat, Joakim
  And{\'e}n, Eugene Belilovsky, et~al.
\newblock Kymatio: Scattering transforms in python.
\newblock \emph{Journal of Machine Learning Research}, 21\penalty0
  (60):\penalty0 1--6, 2020.

\bibitem[Bergen et~al.(2019)Bergen, Chen, and Li]{bergen2019preface}
Karianne~J Bergen, Ting Chen, and Zefeng Li.
\newblock Preface to the focus section on machine learning in seismology.
\newblock \emph{Seismological Research Letters}, 90\penalty0 (2A):\penalty0
  477--480, 2019.

\bibitem[Chao et~al.(2012)Chao, Peng, Wu, Tang, and Lin]{chao2012remote}
K.~Chao, Z.~Peng, C.~Wu, C.C. Tang, and C.H. Lin.
\newblock Remote triggering of non‐volcanic tremor around taiwan.
\newblock \emph{Geophys. J. Int.}, 188:\penalty0 301– 324, 2012.
\newblock \doi{doi:10.1111/j.1365‐246X.2011.05261.x}.

\bibitem[Civilini et~al.(2021)Civilini, Weber, Jiang, Phillips, and
  Pan]{10.1093/gji/ggab083}
F~Civilini, R~C Weber, Z~Jiang, D~Phillips, and W~David Pan.
\newblock {Detecting moonquakes using convolutional neural networks, a
  non-local training set, and transfer learning}.
\newblock \emph{Geophysical Journal International}, 225\penalty0 (3):\penalty0
  2120--2134, 03 2021.
\newblock ISSN 0956-540X.
\newblock \doi{10.1093/gji/ggab083}.
\newblock URL \url{https://doi.org/10.1093/gji/ggab083}.

\bibitem[Gomberg et~al.(2008)Gomberg, Rubinstein, Peng, Creager, Vidale, and
  Bodin]{gomberg2008widespread}
Joan Gomberg, Justin~L Rubinstein, Zhigang Peng, Kenneth~C Creager, John~E
  Vidale, and Paul Bodin.
\newblock Widespread triggering of nonvolcanic tremor in california.
\newblock \emph{Science}, 319\penalty0 (5860):\penalty0 173--173, 2008.

\bibitem[Gutenberg and Richter(1954)]{gutenberg1954seismicity}
B~Gutenberg and CF~Richter.
\newblock Seismicity of the earth 2nd ed., 310, 1954.

\bibitem[Hill and Prejean(2015)]{hill2015dynamic}
David~P. Hill and Stephanie Prejean.
\newblock Dynamic triggering.
\newblock \emph{Treatise on Geophysics}, 4, Earthquake Seismology:\penalty0
  273--304, 2015.
\newblock \doi{doi.org/10.1016/B978-0-444-53802-4.00078-6}.

\bibitem[Kano et~al.(2018)Kano, Aso, Matsuzawa, Ide, Annoura, Arai, Baba,
  Bostock, Chao, Heki, Itaba, Ito, Kamaya, Maeda, Maury, Nakamura, Nishimura,
  Obana, Ohta, Poiata, Rousset, Sugioka, Takagi, Takahashi, Takeo, Tu, Uchida,
  Yamashita, and Obara]{kano2018development}
Masayuki Kano, Naofumi Aso, Takanori Matsuzawa, Satoshi Ide, Satoshi Annoura,
  Ryuta Arai, Satoru Baba, Michael Bostock, Kevin Chao, Kosuke Heki, Satoshi
  Itaba, Yoshihiro Ito, Noriko Kamaya, Takuto Maeda, Julie Maury, Mamoru
  Nakamura, Takuya Nishimura, Koichiro Obana, Kazuaki Ohta, Natalia Poiata,
  Baptiste Rousset, Hiroko Sugioka, Ryota Takagi, Tsutomu Takahashi, Akiko
  Takeo, Yoko Tu, Naoki Uchida, Yusuke Yamashita, and Kazushige Obara.
\newblock {Development of a Slow Earthquake Database}.
\newblock \emph{Seismological Research Letters}, 89\penalty0 (4):\penalty0
  1566--1575, 06 2018.
\newblock ISSN 0895-0695.
\newblock \doi{10.1785/0220180021}.
\newblock URL \url{https://doi.org/10.1785/0220180021}.

\bibitem[Li et~al.(2018)Li, Narvekar, Nakshatra, Raut, Sirkeci, and
  Gao]{li2018seismic}
Wenrui Li, Nishita Narvekar, Nakshatra Nakshatra, Nitisha Raut, Birsen Sirkeci,
  and Jerry Gao.
\newblock Seismic data classification using machine learning.
\newblock In \emph{2018 IEEE Fourth International Conference on Big Data
  Computing Service and Applications (BigDataService)}, pages 56--63. IEEE,
  2018.

\bibitem[Lin et~al.(2018)Lin, Wang, Thiagarajan, Guthrie, and
  Coblentz]{10.1093/gji/ggy385}
Youzuo Lin, Shusen Wang, Jayaraman Thiagarajan, George Guthrie, and David
  Coblentz.
\newblock {Efficient data-driven geologic feature characterization from
  pre-stack seismic measurements using randomized machine learning algorithm}.
\newblock \emph{Geophysical Journal International}, 215\penalty0 (3):\penalty0
  1900--1913, 09 2018.
\newblock ISSN 0956-540X.
\newblock \doi{10.1093/gji/ggy385}.
\newblock URL \url{https://doi.org/10.1093/gji/ggy385}.

\bibitem[Liu et~al.(2020)Liu, Ren, Chen, and Chen]{10.1093/gji/ggaa444}
Xinliang Liu, Tao Ren, Hongfeng Chen, and Yufeng Chen.
\newblock {Classification of tectonic and non-tectonic seismicity based on
  convolutional neural network}.
\newblock \emph{Geophysical Journal International}, 224\penalty0 (1):\penalty0
  191--198, 09 2020.
\newblock ISSN 0956-540X.
\newblock \doi{10.1093/gji/ggaa444}.
\newblock URL \url{https://doi.org/10.1093/gji/ggaa444}.

\bibitem[Liu et~al.(2019)Liu, Yeh, Chen, Chen, Yen, and
  Yen]{liu2019investigation}
Yi-Hung Liu, Ting-Chen Yeh, Kate~Huihsuan Chen, Yaochieh Chen, Yuan-Yi Yen, and
  Horng-Yuan Yen.
\newblock Investigation of single-station classification for short tectonic
  tremor in taiwan.
\newblock \emph{Journal of Geophysical Research: Solid Earth}, 124\penalty0
  (8):\penalty0 8803--8822, 2019.

\bibitem[Meier et~al.(2019)Meier, Ross, Ramachandran, Balakrishna, Nair,
  Kundzicz, Li, Andrews, Hauksson, and Yue]{meier2019reliable}
Men-Andrin Meier, Zachary~E Ross, Anshul Ramachandran, Ashwin Balakrishna,
  Suraj Nair, Peter Kundzicz, Zefeng Li, Jennifer Andrews, Egill Hauksson, and
  Yisong Yue.
\newblock Reliable real-time seismic signal/noise discrimination with machine
  learning.
\newblock \emph{Journal of Geophysical Research: Solid Earth}, 124\penalty0
  (1):\penalty0 788--800, 2019.

\bibitem[Obara(2002)]{obaratremor}
K.~Obara.
\newblock Nonvolcanic deep tremor associated with subduction in southwest
  japan.
\newblock \emph{Science}, 296:\penalty0 1679--1681, 2002.
\newblock \doi{10.1126/science.1070378}.

\bibitem[Oyallon et~al.(2013)Oyallon, Mallat, and Sifre]{oyallon2013generic}
Edouard Oyallon, St{\'e}phane Mallat, and Laurent Sifre.
\newblock Generic deep networks with wavelet scattering.
\newblock \emph{arXiv preprint arXiv:1312.5940}, 2013.

\bibitem[Reynen and Audet(2017{\natexlab{a}})]{10.1093/gji/ggx238}
Andrew Reynen and Pascal Audet.
\newblock {Supervised machine learning on a network scale: application to
  seismic event classification and detection}.
\newblock \emph{Geophysical Journal International}, 210\penalty0 (3):\penalty0
  1394--1409, 05 2017{\natexlab{a}}.
\newblock ISSN 0956-540X.
\newblock \doi{10.1093/gji/ggx238}.
\newblock URL \url{https://doi.org/10.1093/gji/ggx238}.

\bibitem[Reynen and Audet(2017{\natexlab{b}})]{reynen2017supervised}
Andrew Reynen and Pascal Audet.
\newblock Supervised machine learning on a network scale: Application to
  seismic event classification and detection.
\newblock \emph{Geophysical Journal International}, 210\penalty0 (3):\penalty0
  1394--1409, 2017{\natexlab{b}}.

\bibitem[Riggelsen and Ohrnberger(2014)]{riggelsen2014machine}
Carsten Riggelsen and Matthias Ohrnberger.
\newblock A machine learning approach for improving the detection capabilities
  at 3c seismic stations.
\newblock \emph{Pure and Applied Geophysics}, 171\penalty0 (3):\penalty0
  395--411, 2014.

\bibitem[Ross et~al.(2018)Ross, Meier, Hauksson, and
  Heaton]{ross2018generalized}
Zachary~E Ross, Men-Andrin Meier, Egill Hauksson, and Thomas~H Heaton.
\newblock Generalized seismic phase detection with deep learning.
\newblock \emph{Bulletin of the Seismological Society of America}, 108\penalty0
  (5A):\penalty0 2894--2901, 2018.

\bibitem[Rouet-Leduc et~al.(2017)Rouet-Leduc, Hulbert, Lubbers, Barros,
  Humphreys, and Johnson]{rouet2017machine}
Bertrand Rouet-Leduc, Claudia Hulbert, Nicholas Lubbers, Kipton Barros, Colin~J
  Humphreys, and Paul~A Johnson.
\newblock Machine learning predicts laboratory earthquakes.
\newblock \emph{Geophysical Research Letters}, 44\penalty0 (18):\penalty0
  9276--9282, 2017.

\bibitem[Rouet-Leduc et~al.(2018)Rouet-Leduc, Hulbert, and
  Johnson]{rouet2018breaking}
Bertrand Rouet-Leduc, Claudia Hulbert, and Paul~A Johnson.
\newblock Breaking cascadia's silence: Machine learning reveals the constant
  chatter of the megathrust.
\newblock \emph{arXiv preprint arXiv:1805.06689}, 2018.

\bibitem[Ruano et~al.(2014)Ruano, Madureira, Barros, Khosravani, Ruano, and
  Ferreira]{ruano2014seismic}
Ant{\'o}nio~E Ruano, Guilherme Madureira, Ozias Barros, Hamid~Reza Khosravani,
  M~Gra{\c{c}}a Ruano, and Pedro~M Ferreira.
\newblock Seismic detection using support vector machines.
\newblock \emph{Neurocomputing}, 135:\penalty0 273--283, 2014.

\bibitem[Seydoux et~al.(2020)Seydoux, Balestriero, Poli, De~Hoop, Campillo, and
  Baraniuk]{seydoux2020clustering}
L{\'e}onard Seydoux, Randall Balestriero, Piero Poli, Maarten De~Hoop, Michel
  Campillo, and Richard Baraniuk.
\newblock Clustering earthquake signals and background noises in continuous
  seismic data with unsupervised deep learning.
\newblock \emph{Nature communications}, 11\penalty0 (1):\penalty0 1--12, 2020.

\bibitem[Tang et~al.(2020{\natexlab{a}})Tang, R{\"o}sler, Nelson, Thompson,
  van~der Lee, Chao, and Paulsen]{tang2020citizen}
Vivian Tang, Boris R{\"o}sler, Jordan Nelson, JaCoya Thompson, Suzan van~der
  Lee, Kevin Chao, and Michelle Paulsen.
\newblock Citizen scientists help detect and classify dynamically triggered
  seismic activity in alaska.
\newblock \emph{Frontiers in Earth Science}, 8:\penalty0 321,
  2020{\natexlab{a}}.

\bibitem[Tang et~al.(2020{\natexlab{b}})Tang, Seetharaman, Chao, Pardo, and
  van~der Lee]{tang2020automating}
Vivian Tang, Prem Seetharaman, Kevin Chao, Bryan~A Pardo, and Suzan van~der
  Lee.
\newblock Automating the detection of dynamically triggered earthquakes via a
  deep metric learning algorithm.
\newblock \emph{Seismological Research Letters}, 91\penalty0 (2A):\penalty0
  901--912, 2020{\natexlab{b}}.

\bibitem[Tang et~al.(2021)Tang, Chao, and van~der Lee]{tang2021}
Vivian Tang, Kevin Chao, and Suzan van~der Lee.
\newblock {Detections of Directional Dynamic Triggering in Intraplate Regions
  of the United States}.
\newblock \emph{Bulletin of the Seismological Society of America}, 04 2021.
\newblock ISSN 0037-1106.
\newblock \doi{10.1785/0120200352}.
\newblock URL \url{https://doi.org/10.1785/0120200352}.

\bibitem[Velasco et~al.(2008)Velasco, Hernandez, Parsons, and
  Pankow]{velasco2008global}
Aaron~A Velasco, Stephen Hernandez, TOM Parsons, and Kris Pankow.
\newblock Global ubiquity of dynamic earthquake triggering.
\newblock \emph{Nature geoscience}, 1\penalty0 (6):\penalty0 375--379, 2008.

\bibitem[Wiszniowski et~al.(2021)Wiszniowski, Plesiewicz, and
  Lizurek]{wiszniowski2021machine}
Jan Wiszniowski, Beata Plesiewicz, and Grzegorz Lizurek.
\newblock Machine learning applied to anthropogenic seismic events detection in
  lai chau reservoir area, vietnam.
\newblock \emph{Computers \& Geosciences}, 146:\penalty0 104628, 2021.

\bibitem[Yun et~al.(2021)Yun, Yang, and Zhou]{yun2021dyntripy}
Naidan Yun, Hongfeng Yang, and Shiyong Zhou.
\newblock Dyntripy: A python package for detecting dynamic earthquake
  triggering signals.
\newblock \emph{Seismological Society of America}, 92\penalty0 (1):\penalty0
  543--554, 2021.

\end{thebibliography}

\section*{Appendix A: Data}\label{App A}
\subsection*{Clean Data}
\begin{enumerate}
    \item A set of 1000 s-long waveforms with confirmed PT tremor signals, labeled by the seismologists among us. The waveforms were resampled at 20 samples per second and band-pass filtered between 2 and 5 Hz. This data set is subdivided as follows:
    \begin{enumerate}
        \item  Waveforms recorded by seismic stations in Taiwan or Japan, trimmed around surface waves from 6 large-M earthquakes in the eastern hemisphere (the Great Tohoku Earthquake, 1 more earthquake from Hokkaido, Japan, 1 from Qinghai, China, and 4 from Sumatra, Indonesia). These are the positive examples.
        \item Waveforms recorded by seismic stations in Taiwan, Malaysia, Australia, and of the Global Seismographic Network (GSN), selected for having no significant signals from earthquakes or otherwise. These are the negative examples.
    \end{enumerate}
    \item A set of 1000 s-long waveforms with confirmed PT local earthquake signals, labeled by the seismologists among us. All data was band-pass filtered between 2 and 8 Hz.  
    \begin{enumerate}
        \item Waveforms recorded by seismic stations from USArray in the USA and the Hi-CLIMB array in Tibet, trimmed around surface waves from the 2010 M8.8 Maule Earthquake and 7 additional large-M earthquakes in the eastern hemisphere (1 from China, 1 from Japan, and 5 from Sumatra, Indonesia). Waveforms that showed signals from local earthquakes were labeled as positive.
        \item  The waveforms from this set (2a) that were not labeled positive - they were labeled as negative examples. 
        \item Additional waveforms recorded by the Hi-CLIMB array from 10 random local earthquakes with M$<$3.6. This auxiliary data set was used to expand the set of positive examples. An corresponding number of negative examples recorded by the same array was added to the negative examples. 
    \end{enumerate}

\end{enumerate}

\begin{figure}
 \centering 
  \includegraphics[width=\linewidth]{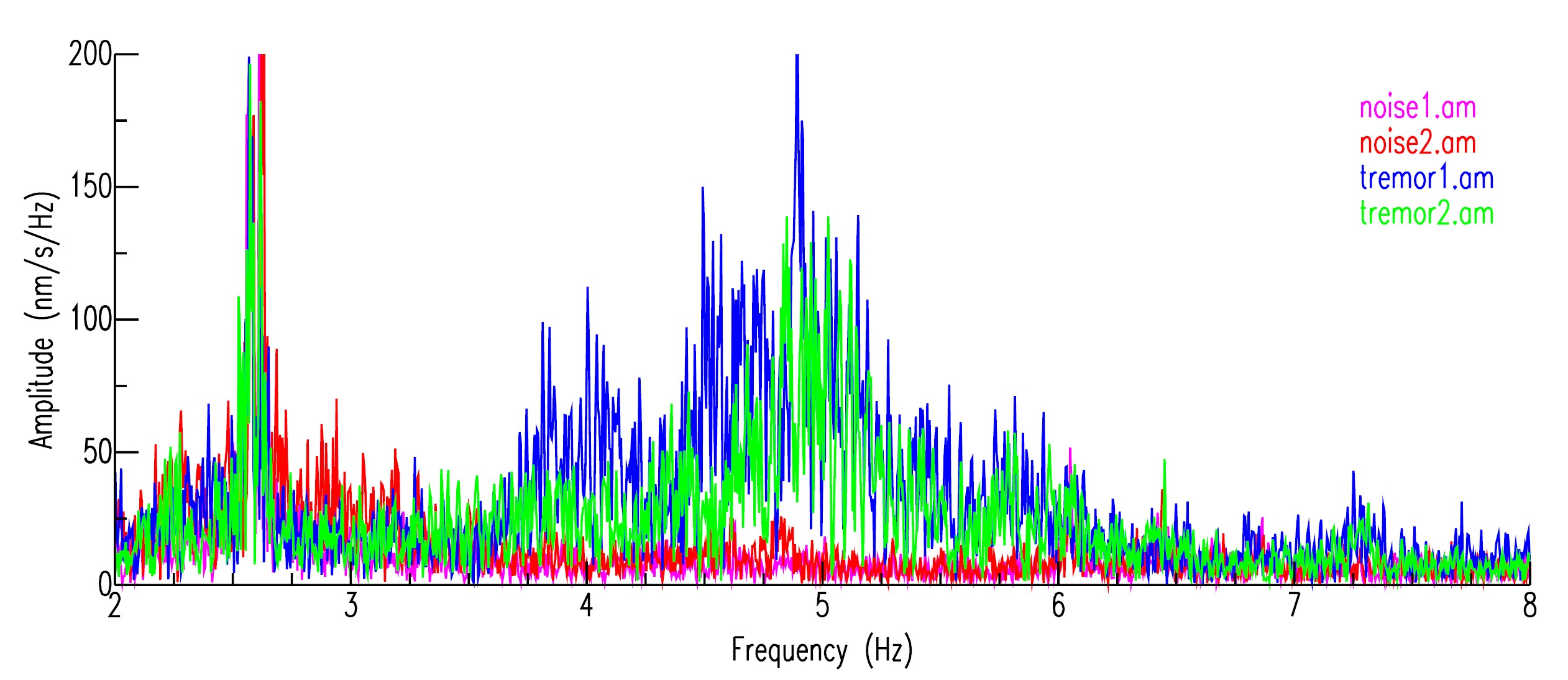}
  \caption{Amplitude spectra for four 200-s subsets of a 2000-s wave train labeled as tremor and misclassified as noise. Two subsets (red and magenta) represent noise and two (blue and green) contain a tremor-like signal.}
  \label{power_comp}
\end{figure}

\subsection*{Data Distribution}
\begin{itemize}
    \item In the first experiment we used 551 earthquake samples, 570 noise samples and 39 tremor samples from the clean data. 
    \item In the second experiment (clean + gold users data), we had the following data distribution: 1031 earthquake samples, 1014 noise samples and 48 tremor samples. 
    \item In the final experiment (which additionally includes the chosen volunteer's data) had the following data distribution: 3013 earthquake samples, 2436 noise samples, 203 tremor samples. 
\end{itemize}

\section*{Appendix B: Model Analysis}\label{App B}

To gain further insights into how the WavImg model classifies, we looked at select examples. One example is where the ML algorithm classified a wave train with tremor as a noise wave train. This could be the result of 
\begin{itemize}
\item There being two separate bursts of tremor,
\item Both tremor signals being relatively short in duration,
\item The tremor signals being weak,
\item The tremor signals  sounding different from more typical tremor signals
\item The presence of non-stationary noise signals. 
\end{itemize}

Extracting two 200-s tremor signals and two 200-s stationary noise signals from the 2000-s wave train reveals that the the weak tremor signals have some additional power between 3 and 6 Hz, compared to the noise (Figure \ref{power_comp}). This is not entirely characteristic but still consistent for tremor signals. Although the wave train was likely labeled correctly, the wave train is not a role model for its class and hence may have confused the ML algorithm. In one other case labeled as tremor and classified as earthquake, the tremor signal was so brief that is easy to mistake for an earthquake signal. In another case, strongly peaked signals elsewhere in the wave train might have distracted the ML from the tremor signal. 

The type of waveform data used in our study contains a wild variety of noise signals, for which we did not designate a single class. However such noise signals can interfere with the ability of volunteers, and sometimes experts to correctly label wave trains. In at least 5 cases, wave trains with noise signals, labeled as noise, were misclassified as earthquakes.

Several other cases of misclassification by the ML algorithm can be traced to a mislabeling of the original data. In three cases of wave trains with noise signals mislabeled as tremor, the ML algorithm classified the wave trains as earthquakes. The ML algorithm also classified a case of mislabeled tremor correctly as noise. In at least 4 cases, the ML algorithm correctly classified wave trains as noise, while they were labeled as earthquakes. In these four cases, listening to and viewing spectral properties of the wave trains confirmed in hindsight that these signals should have indeed been labeled as noise.

\end{document}